# Interaction of hearing aids with self-motion and the influence of hearing impairment

Maartje M. E. Hendrikse[1*], Theda Eichler[1], Giso Grimm[1], Volker Hohmann[1]

[1]Auditory Signal Processing and Cluster of Excellence "Hearing4all",
Department of Medical Physics and Acoustics, University of Oldenburg, Oldenburg, Germany.

*Corresponding author: Maartje Hendrikse, maartje.hendrikse@uol.de

## Abstract

When listening to a sound source in everyday-life situations, typical movement behavior can lead to a mismatch between the direction of the head and the direction of interest. This could reduce the performance of directional algorithms, as was shown in previous work for head movements of normal-hearing listeners. However, the movement behavior of hearing-impaired listeners and hearing aid users might be different, and if hearing aid users adapt their self-motion because of the directional algorithm, its performance might increase. In this work we therefore investigated the influence of hearing impairment on self-motion, and the interaction of hearing aids with self-motion. In order to do this, the self-motion of three hearing-impaired (HI) participant groups, aided with an adaptive differential microphone (ADM), aided without ADM, and unaided, was compared, also to previously measured self-motion data from younger and older normal-hearing (NH) participants. The self-motion was measured in virtual audiovisual environments (VEs) in the laboratory. Furthermore, the signal-to-noise ratios (SNRs) and SNR improvement of the ADM resulting from the head movements of the participants were estimated with acoustic simulations. A strong effect of hearing impairment on self-motion was found, which led to an overall increase in estimated SNR of 0.8 dB for the HI participants compared to the NH participants, and differences in estimated SNR improvement of the ADM. However, the self-motion of the HI participants aided with ADM and the other HI participants was very similar, indicating that they did not adapt their self-motion because of the ADM.

**Key words: hearing aids, adaptive differential microphone, self-motion**

# Introduction

Typical movement behavior in everyday-life situations can lead to a mismatch between the direction of the head and the direction of interest when listening to a sound source. This can happen because people do not always look in the direction of interest, for example when the sound source is not visible or when they get distracted, and because they do part of the movement with their eyes (Hendrikse, Llorach, Hohmann, & Grimm, 2019a). Moreover, in multi-talker conversations it is natural to sometimes look away from the active speaker for social reasons (Vertegaal, Slagter, Van Der Veer, & Nijholt, 2001). For directional hearing aid algorithms this would mean that they are not always facing in the optimal direction, which could reduce their performance. This was described by Hendrikse, Grimm, & Hohmann (2020) as the effect of misalignment. They showed that natural variances in head movements of normal-hearing listeners can indeed result in a different performance of directional algorithms across individual listeners because of misalignment, and that this is a large effect. Ricketts (2000), Abdipour et al. (2015), and Boyd et al. (2013) have also shown an effect of head movement on directional algorithms. It is not yet known whether the movement behavior of hearing-impaired listeners and hearing aid users is different in everyday-life listening situations. There are studies that investigated the influence of hearing impairment and directional microphones on orienting behavior (Brimijoin, McShefferty, & Akeroyd, 2010; Brimijoin, Whitmer, McShefferty, & Akeroyd, 2014), showing that hearing impairment and directional microphones are associated with an increased complexity of orienting behavior towards auditory-only targets. So, there could also be changes in everyday-life listening behavior associated with hearing loss and hearing aid use. Perhaps hearing aid users adapt their head movement behavior when using a directional algorithm, which might lead to a better performance of the algorithms.

In this study we therefore investigate how hearing loss and hearing aids with and without a directional algorithm affect self-motion. We talk about self-motion, i.e. the rotation and translation of the head and torso along with the movement of the eyes, rather than movement behavior here, because we cannot measure psychological aspects such as the intent of the motion due to the limited ecological validity of the laboratory environment (Grimm, Hendrikse, & Hohmann, 2020; Hohmann, Paluch, Krueger, Meis, & Grimm, 2020). To identify whether potential changes in self-motion are relevant in the context of hearing research, we also investigate their influence on the signal-to-noise ratio (SNR) and SNR improvement provided by the directional algorithm.

To do this, older hearing-impaired listeners with and without hearing aid experience were recruited. The participants listened to the virtual audiovisual environments (living room, lecture hall, cafeteria, street and train station) from (Hendrikse et al., 2019a), with some small adaptations to adjust the difficulty. Meanwhile their head, eye and body movements were measured. The hearing loss of the experienced hearing aid users was compensated. For half of the experienced hearing aid users, the sound was processed with a directional algorithm in addition to the hearing loss compensation. An adaptive differential microphone ("ADM"), after Elko & Pong (Elko & Pong, 1995), was chosen as directional algorithm, because adaptive differential microphones are often used in commercial hearing aids. Binaural rendering with head-tracking controlled simulation of self-motion was used to render the sound. The self-motion of the participants was analyzed and compared to that of the normal-hearing participants measured in previous work (Hendrikse et al., 2019a). Moreover, the influence of the self-motion of each participant group on the SNR and the SNR improvement by the ADM was analyzed using acoustic simulations. Estimating the SNR improvement of the ADM also for the participant groups who did not actually use the ADM allowed us to determine whether the aided participants with ADM adapted their self-motion to improve its performance.

It has been shown that hearing-impaired listeners with asymmetric hearing loss successfully make use of head movements to increase the level of the target (Brimijoin, McShefferty, & Akeroyd, 2012). This strategy might not maximize the SNR, but it can help to improve it. Grange & Culling showed that young normal-hearing listeners had difficulty to find a beneficial head orientation (Grange & Culling, 2016). Both studies from Brimijoin et al. and Grange & Culling were performed without visual cues. When visual cues are present it has been shown that participants usually look at the target speaker (Hendrikse, Llorach, Grimm, & Hohmann, 2018; Hendrikse et al., 2019a), but differences were found between younger and older normal-hearing participants in the relative amount of movement that was done with the head and the eyes (head-eye ratio). Therefore, we hypothesize that hearing impaired listeners and hearing aid users, like normal-hearing listeners, look at the target speaker. We also expect that hearing impaired listeners and hearing aid users orient their head differently, resulting in an increased SNR compared to normal-hearing listeners. If this is the case, the head-eye ratio will differ from normal-hearing listeners. In line with this, we expect that the participants using the ADM adapt their head orientation to increase the SNR improvement of the ADM.

## Method

### Participants

A total of 30 older (50-80 years, mean age 72.2 years) hearing-impaired (HI) participants was included in this study. The participants were native German speakers and did not suffer from neck- or back-problems, dizziness or epilepsy. All participants had a moderate, symmetric, sensorineural hearing loss resembling the N3 standard audiogram according to Bisgaard et al. (Bisgaard, Vlaming, & Dahlquist, 2010). The participants were divided into three groups: unaided, aided with ADM and aided without ADM. In the unaided group, nine participants were included who did not wear a hearing aid in everyday life, or had a hearing aid for less than two years. The remaining 21 participants had a hearing aid for more than two years (mean 9.7 years), and wore the hearing aid on average 10.7 hours a day. These 21 hearing aid users were randomly distributed over the aided without ADM (10 participants) and aided with ADM (11 participants) groups.

### Environments and tasks

The virtual audiovisual environments (VEs) from Hendrikse et al. (Hendrikse et al., 2019a; Hendrikse, Llorach, Hohmann, & Grimm, 2019b) were used in this study, with a few small adaptations. These VEs represented: listening to the news in a living room, listening to a lecture in a lecture hall, listening to a multi-person conversation in a cafeteria, listening to announcements in a train station, listening to a multi-person conversation at a bus stop on a street (hereafter: street$_{active}$), and waiting at a cross-section (hereafter: street$_{passive}$). The cafeteria VE was presented twice. During the first presentation (hereafter: cafeteria$_{listeningonly}$ VE) the task of the participant was to follow the conversation between the animated persons at the table. The second presentation (hereafter: cafeteria$_{dualtask}$ VE) tested the hand-eye coordination of the test persons in addition. The subjects were asked to follow the conversation and at the same time, on the Purdue Pegboard (Tiffin & Asher, 1948), insert one pin after the other into the holes. A more detailed description of the VEs, including presentation levels, can be found in Appendix 1, pictures are shown in Figure 1. Hendrikse et al. concluded that these VEs were sufficiently realistic, but some suggestions for improvement were made based on the comments of the participants. Most importantly, the conversations in the cafeteria VE were seen as unrealistic by the participants, because no Lombard speech was included in these conversations, making them too hard to understand in such a noisy environment despite realistic SNR. Therefore, new conversations were recorded and implemented including Lombard speech, with three actors (Gerken, Hendrikse, Grimm, & Hohmann, 2020). To achieve this, the three actors in the conversations were listening to cafeteria noise via headphones while their speech was recorded.

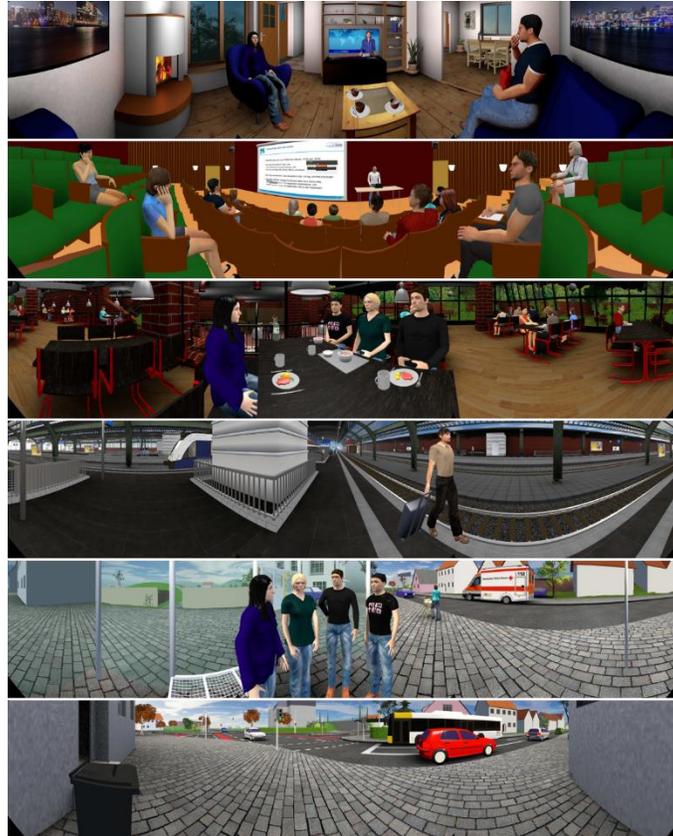

**Figure 1:** Pictures of the virtual audiovisual environments (VEs) from Hendrikse et al. (Hendrikse et al., 2019a, 2019b). From top to bottom: living room, lecture hall, cafeteria, train station, street$_{active}$, and street$_{passive}$.

When recording the self-motion of the participants, it was important that they made an effort to understand the target and did not "give up". Therefore, the SNR of the VEs was adjusted individually to ensure a better level of difficulty. For the individual adjustment, the 70% speech reception threshold (SRT70) of the participants was measured with the Oldenburg matrix sentence test (OLSA; Wagener, Brand, & Kollmeier, 1999) in the diffuse background noise of the cafeteria. The mean SRT70 of 10 NH participants was subtracted from the individual SRT70 and the overall noise level reduced by this difference. This way, the noise level was on average reduced by 6.8 ± 2.5 dB for the HI participants. Three of the VEs, living room, lecture hall and cafeteria, were presented while the participants were seated on a chair and the other two VEs, train station and street, were presented while the participants were standing. In each VE a target was defined that they had to attend to. Before the actual measurement, the participants could listen to each VE (without target) for about 30 seconds to get familiar with the environment. After the presentation of each VE, the participants had to fill in a content-based multiple-choice questionnaire to make sure they were paying attention to the target.

## Setup

This section explains how the virtual audiovisual environments were presented to the participants, and how the self-motion was measured. The measurements were conducted in the audio lab in Oldenburg, for illustrations see Hendrikse et al. (2018). The setup was placed in a sound-treated room, and the equipment was attached to a cloth-covered metal frame, with a diameter of 3.5 m, that reduced environmental sounds, light and room reflections.

### Visual presentation

The visual stimuli were projected onto a cylindrical, acoustically transparent, screen using three projectors (NEC U321H). The combined field of view was about 300°. A graphics card (Nvidia Quadro m5000) performed the warping necessary for projecting onto a cylindrical screen. This process was manually calibrated. The virtual visual environments had been created in Blender (version 2.78a, Roosendaal, 1995) and were rendered using the built-in game engine of this software package. A simulation of movement parallax was added to increase the presence and involvement of the participants. This changed the visual perspective according to the small sways and slight translations the participants made. The position of the virtual camera and the virtual listening position (see next section) was displaced by half of the physical displacement of the head to account also for the displacement relative to the projection screen and loudspeakers, that is, a head translation of 10 cm would be equivalent to a camera/listening position translation of 5 cm in the virtual environments.

### Audio presentation

The virtual acoustic environments had been created using the software package TASCAR (version 0.202-0.204; Grimm, Luberadzka, & Hohmann, 2019). The overall noise level in all VEs was adjusted individually according to the SRT70 scores of the participants as described in the 'Environments and tasks' section, for presentation levels see Appendix 1. The signals were rendered for a loudspeaker layout of 29 loudspeakers. The loudspeakers were arranged in a 16-loudspeaker horizontal ring array at ear level (first loudspeaker at 11.25° from frontal direction, with 22.5° spacing), two 6-loudspeaker ring arrays at +45° and -45° elevation (first loudspeakers at 0° and 30° azimuth from frontal direction, respectively, with 60° spacing), and one pole loudspeaker (90° elevation, directly above the participant). However, in the current study the signals were not played back over loudspeakers, but over headphones (Sennheiser HDA 200). To achieve this, the rendered signals were convolved with previously measured head- and hearing-aid-related impulse responses (HARIRs; in-ear, plus three hearing aid microphone positions on each side) for the mentioned loudspeaker layout. The impulse responses were recorded as described in Grimm, Schwarte et al. (2020). A preliminary version of the recordings by Grimm, Schwarte, et al. (2020) was used during the measurements in this study, which contained an error in the channel order for the upper and lower loudspeaker rings. The influence of this error on self-motion was assessed in the additional experiment (Appendix 2), by comparing to the final HARIRs from Grimm, Schwarte, et al. (2020). The effect of the error on self-motion was found to be negligible. After convolution, the resulting signals for the front hearing aid microphones were played back over headphones for the unaided group of participants. The HARIRs from the front hearing aid microphones were chosen instead of the in-ear signals to make sure a difference in self-motion would be due to the hearing loss compensation. For the aided group without ADM also the HARIRs from the front hearing aid microphones were used, and hearing loss compensation was applied. For the aided group with ADM, the signals resulting from convolution with HARIRs from front and rear hearing aid microphones were used as input for the ADM, and hearing loss compensation was applied to the ADM output. Moreover, the head translations in the horizontal plane and the head yaw measured by the head movement sensor were sent to TASCAR, and the position of the acoustic receiver adjusted accordingly, so that the presented audio changed according to the head movements the participants made (binaural rendering with head-tracking controlled simulation of self-motion). The audio was processed with a block size of 1024 samples.

### Hearing loss compensation and ADM

The hearing loss compensation and ADM were applied using the open master hearing-aid software (openMHA version 4.9.0, Herzke, Kayser, Loshaj, Grimm, & Hohmann, 2017). The hearing loss compensation was performed based on the measured audiogram of the participants, using the CR2 NAL-RP fitting rule. CR2 NAL-RP is based on the linear NAL-RP fitting rule developed by the National Acoustics Laboratories (Byrne, Parkinson, & Newall, 1990). In addition, it adds a compression of 1:2, with a knee point at the narrow-band levels corresponding to 65 dB SPL LTASS, and a noise gate to make sure soft sounds are not maximally amplified. The implementation of the ADM in the openMHA is based on the algorithm described by Elko & Pong (1995). The signal was processed with the ADM before applying the hearing loss compensation.

### Sensors

Six infrared cameras (Qualisys Miqus M3) were used to measure the head and torso movements. These tracked the subjects' movements via highly reflective markers attached to the headphones (head movement), and the right and left shoulder pads of a vest (torso movement). The three axes of rotation were recorded: roll, pitch and yaw, as well as the translation on these axes. The motion was recorded by the motion tracking software Qualisys (version: 2019.3.4780) with a sampling frequency of 200 Hz.

The eye movement was measured by a custom-made wireless electrooculogram sensor (EOG). One electrode was placed on the temple next to each eye. The measured signal was sent via WLAN with a sampling rate of 33 Hz and a resolution of 16 bits. Thus the eye movement in horizontal plane could be measured with an accuracy of approximately ±10°. The drift in the eye movement data was approximated by linearly extrapolating the data and then smoothing with a moving-average filter with a length of 500 samples. This approximated drift was subtracted from the eye movement data.

The calibration of the sensors was done with a cross projected onto the screen. To align the head movement sensor with the head direction, the participants were asked to adjust the markers on the headphones until the cross was placed as accurately as possible in the direction of their nose. To calibrate the EOG sensor, the cross was displaced 5-30° to the left or right. The participants were asked to follow the cross with their eyes, accidental head movements were compensated for. After switching between the standing/sitting environment, calibration was repeated. To obtain the gaze trajectories, the head- and eye-movement trajectories were resampled to the same time line with a 120-Hz sampling rate and summed. The TASCAR and LabStreamingLayer (Medine, 2016) software packages were used for time synchronization and data logging.

## Analyses

### Analysis of self-motion

To analyze how hearing impairment and hearing aids with and without a directional algorithm affect self-motion, the self-motion of the three HI participant groups, unaided, aided without ADM and aided with ADM, is compared. The movement measures from Hendrikse et al. (2019a) are calculated. These include: the standard deviation of the gaze trajectories (GazeStd); the standard deviation of the head trajectories (HeadStd); the mean speed of the gaze trajectories (GazeSpeedMean); the mean speed of the head trajectories (HeadSpeedMean); the number of gaze jumps, normalized by the duration of the VE (NGazeJumps); the absolute head angle relative to torso over the absolute gaze angle relative to torso (HeadGazeRatio). Furthermore, the root-mean-squared (RMS) error between the direction of the target and the head (TargetHeadRMS) or gaze direction (TargetGazeRMS) was calculated to check how accurately the participants oriented towards the target. In the VEs with multi-talker conversations, the direction of the active speaker at each time point was taken as target direction, and in the lecture hall the direction of the lecturer. In the train station and street$_{passive}$ VEs no target direction could be defined, so the TargetHeadRMS and TargetGazeRMS were not calculated in these VEs.

### Acoustic simulations

The acoustic simulation method from Hendrikse, Schwarte, et al. (2020) was used to estimate the SNRs resulting from the individual head movement trajectories of the HI participants in the VEs. In the acoustic simulations, the rendered signals for the 29-loudspeaker layout as described in the 'Audio presentation' section were effectively rotated and shifted according to the measured head yaw and head translations of the participants. This resulted again in 29-channel loudspeaker signals, now including the movement. Then, the signals were convolved with the final HARIRs from Grimm, Schwarte, et al. (2020), resulting in hearing aid microphone signals. The acoustic simulations were done separately for the target and noise signals, so that the SNR could be calculated. From the hearing aid microphone signals, the better-ear SNR was calculated using segmental SNR after Quackenbush et al. (1988), with 200-ms non-overlapping windows. By comparing these SNRs between participant groups, differences might become apparent that were not visible in the analysis of self-motion, and it can be analyzed whether differences are relevant for assessing acoustic communication ability. As input for the acoustic simulations, the same SNR was used for all HI participants, namely the SNR resulting from the average noise level reduction of 6.8 dB, so that the results could be compared between participants.

In addition, the hearing aid microphone signals resulting from the acoustic simulations were processed with the ADM using the openMHA. To be able to calculate the SNR after processing with the ADM, the method described by Hagerman & Olofsson (2004) was applied. The better-ear SNR was calculated from the processed signals as described in the previous paragraph. It was analyzed how the ADM would have improved the SNR for the different head movement trajectories by taking the difference between the SNRs of the hearing aid microphone signals processed with the ADM and the unprocessed signals for each head movement trajectory (SNRimprovementADM). Because the acoustic simulations were done separately from the recording of the self-motion, this could be done for the head movement trajectories of all participants, regardless of their aided condition when recording the self-motion. Estimating the SNR improvement of the ADM also for the participant groups who did not actually use the ADM allowed us to determine whether the aided participants with ADM adapted their self-motion to improve its performance. Since no target source was defined in the street$_{passive}$ VE, the SNR measures could not be calculated there.

### Data from previous study

The previously recorded self-motion data of 21 younger normal-hearing (NH) and 19 older NH participants was also included in the analyses, and is available online (Hendrikse, Llorach, Hohmann, & Grimm, 2019c). The self-motion of the NH participants was also measured in the previously-mentioned VEs, but at a different SNR than the self-motion of the HI participants. Only the cafeteria VEs were different, because the original target conversations with four talkers from the previous study were exchanged for the Lombard conversations with three talkers in the current study. Another difference is that the VEs were presented to the NH participants via loudspeakers, and to the HI participants over headphones. The influence of wearing headphones on self-motion was assessed in an additional experiment by comparing to presentation via loudspeakers (Appendix 2) to check whether a comparison between the NH and HI self-motion data is valid. No difference in self-motion was found between headphone and loudspeaker presentation. The measures from the 'Analysis of self-motion' section were also calculated for the self-motion data of the NH participants.

In Hendrikse, Grimm, et al. (2020), the NH self-motion data was used in acoustic simulations to estimate the SNRs resulting from the individual head movement trajectories. The acoustic simulation method was revised to make it more accessible, which did not change the outcome, and the scripts were published in a database (Hendrikse, Schwarte, Grimm, & Hohmann, 2020). The estimated SNRs for the NH self-motion data from this database were also used in the current study for comparison to the estimated SNRs for the HI self-motion data. However, because the average noise level reduction of 6.8 dB relative to the NH participants was applied in the acoustic simulations for the HI participants, and the target conversations in the cafeteria VEs were different, the estimated SNRs resulting from the self-motion of the HI and NH participants could not be compared directly. To allow for a comparison, the SNRs without movement (always facing frontal direction) for the VEs from the current and previous study were taken as a reference (Table 1), and the SNRs including the head movement of the HI and NH participants were calculated relative to this reference (SNRrelative). The SNRimprovementADM data for the NH participants was also taken from the database and included in the comparison. A different SNR could influence the performance of the ADM, being a nonlinear algorithm, so a comparison between HI and NH data should be taken with care.

**Table 1: Reference SNRs without movement (always facing frontal direction) for each participant group.**

| Participants | cafeteria$_{dualtask}$ | cafeteria$_{listeningonly}$ | lecture hall | living room | train station | street$_{active}$ |
|---|---|---|---|---|---|---|
| NH* | -2.1 dB | -5.0 dB | 4.0 dB | 8.4 dB | -8.5 dB | 2.3 dB |
| HI | 5.1 dB | 3.6 dB | 11.7 dB | 13.3 dB | -0.4 dB | 10.5 dB |

*Data from previous study.

### Statistics

A multiway ANOVA was done for the movement measures GazeStd, HeadStd, GazeSpeedMean, HeadSpeedMean, NGazeJumps and HeadGazeRatio. The participant group (unaided, aided without ADM, aided with ADM, and the NH younger and NH older participants from previous work) was included as a between-participants factor. In previous work a significant effect of the different VEs on the self-motion was found, so the environment was included as a within-participant factor, but only to check for interaction effects with the participant group. Similarly, two separate multiway ANOVAs were done for the TargetHeadRMS and TargetGazeRMS measures, and the SNR measures SNRrelative and SNRimprovementADM, but without the train station and/or street$_{passive}$ VEs. To correct for multiple comparisons (Cramer et al., 2016), the false discovery rate was controlled with the Benjamini-Hochberg procedure (Benjamini & Hochberg, 1995) for the grouped results of all ANOVAs.

# Results

## Analysis of Self-motion

The main effects of the multiway ANOVAs for the movement measures are shown in Table 2. The outcomes show that there is a strong effect of the participant group on the HeadGazeRatio, and a weaker effect on the GazeStd. Moreover, there are weak interaction effects between the participant group and environment for all measures except HeadSpeedMean. In the following, the measures that have a significant effect of the participant group are analyzed in more detail. Because all these measures also had significant interaction effects, pairwise comparisons (with Bonferroni correction) were done to check for differences between participant groups in each VE.

**Table 2: Main effects of multiway ANOVAs for the movement measures.**

| Factor | Measure | F-value | p | p$_{adj}$ | Effect Size |
|---|---|---|---|---|---|
| group | GazeStd | F(4,63)=3.2 | .018 | .030* | 0.18 |
| | HeadStd | F(4,63)=2.2 | .081 | .090 | 0.13 |
| | GazeSpeedMean | F(4,63)=2.6 | .045 | .056 | 0.15 |
| | HeadSpeedMean | F(4,63)=2.1 | .089 | .093 | 0.13 |
| | NGazeJumps | F(4,63)=2.4 | .062 | .078 | 0.14 |
| | HeadGazeRatio | F(4,63)=138.6 | <.001 | <.001*** | 0.91 |
| | TargetGazeRMS | F(4,63)=4.3 | .036 | .052 | 0.22 |
| | TargetHeadRMS | F(4,63)=2.7 | .612 | .612 | 0.15 |
| group* environment | GazeStd | F(15.8,229.5)=3.4 | <.001 | <.001*** | 0.19 |
| | HeadStd | F(12.5,180.6)=2.1 | .019 | .030* | 0.13 |

|   | GazeSpeedMean | F(16.3,236.0)=3.7 | <.001 | <.001*** | 0.20 |
|   | HeadSpeedMean | F(10.5,152.0)=2.0 | .036 | .052 | 0.12 |
|   | NGazeJumps | F(19.3,280.3)=4.0 | <.001 | <.001*** | 0.22 |
|   | HeadGazeRatio | F(17.9,259.7)=2.3 | .003 | .005** | 0.14 |
|   | TargetGazeRMS | F(11.1,175.6)=3.2 | <.001 | .001** | 0.17 |
|   | TargetHeadRMS | F(9.7,152.8)=3.8 | <.001 | <.001*** | 0.20 |

*Note.* The $p_{adj}$-column lists p-values adjusted for multiple comparisons (Benjamini-Hochberg procedure). Significances are indicated at the *.05 level **.01 level ***.001 level.

To look into the significant effect of the participant group on the HeadGazeRatio, the HeadGazeRatio is shown in Figure 2 per participant group and VE, and significant differences according to pairwise comparisons are indicated. Overall, the HI participants have a significantly higher HeadGazeRatio than the NH participants (p<.001), indicating that they did relatively more of the movement with their head. In fact, the HeadGazeRatio of the HI participants is almost 1, so they did almost all of the movement with their head. This can be partly attributed to age, because the NH older participants had a significantly higher HeadGazeRatio than the NH younger participants in most of the VEs, but seems to be mainly due to hearing impairment. Although the HI participants did more of the movement with their head than the NH participants, this did not result in a more accurate orientation of their head towards the target direction (no significant participant group differences in TargetHeadRMS).

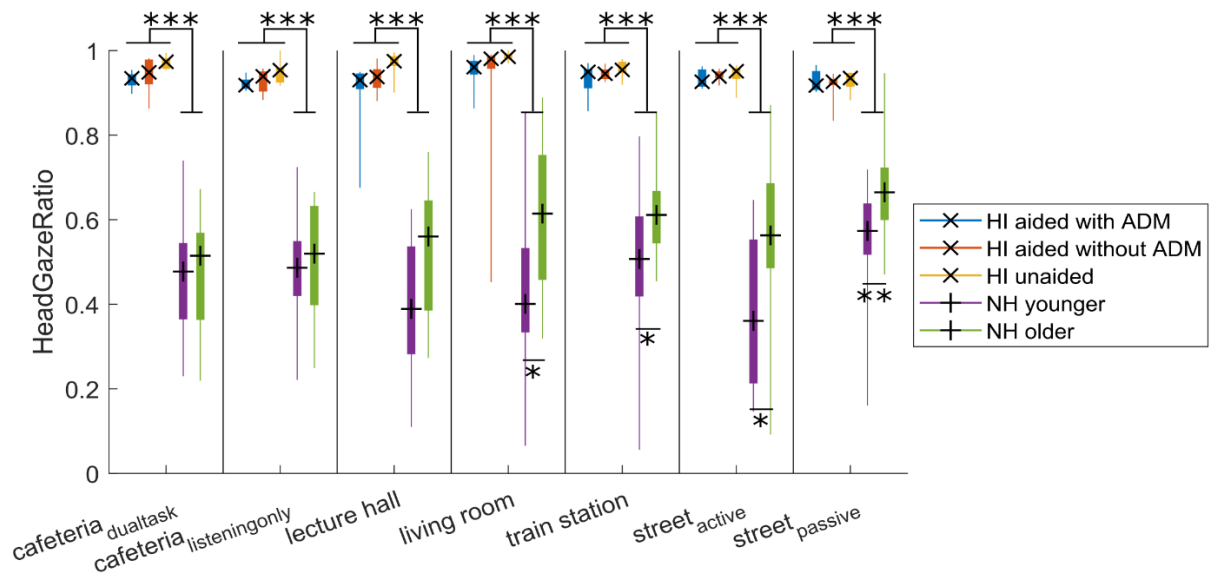

**Figure 2: Boxplots of the HeadGazeRatio (1 = all head movement, 0 = all eye movement) for all participant groups in all VEs. Significant differences are indicated at the *.05 level, **.01 level, ***.001 level. Boxplots show the range (thin line), 25th and 75th percentiles (thick line) and the median (cross).**

The same was done for the GazeStd (Figure 3). Overall, the unaided participants had a significantly lower GazeStd (p=0.046) than the NH older participants, which can be seen in some, but not all, of the VEs in Figure 3. Thus, the unaided participants made slightly less gaze movements than the NH older participants. This effect can be attributed to hearing impairment, but seems to disappear when wearing hearing aids. Only in the cafeteria$_{listeningonly}$ VE, all HI participants had a significantly larger GazeStd than the NH participants.

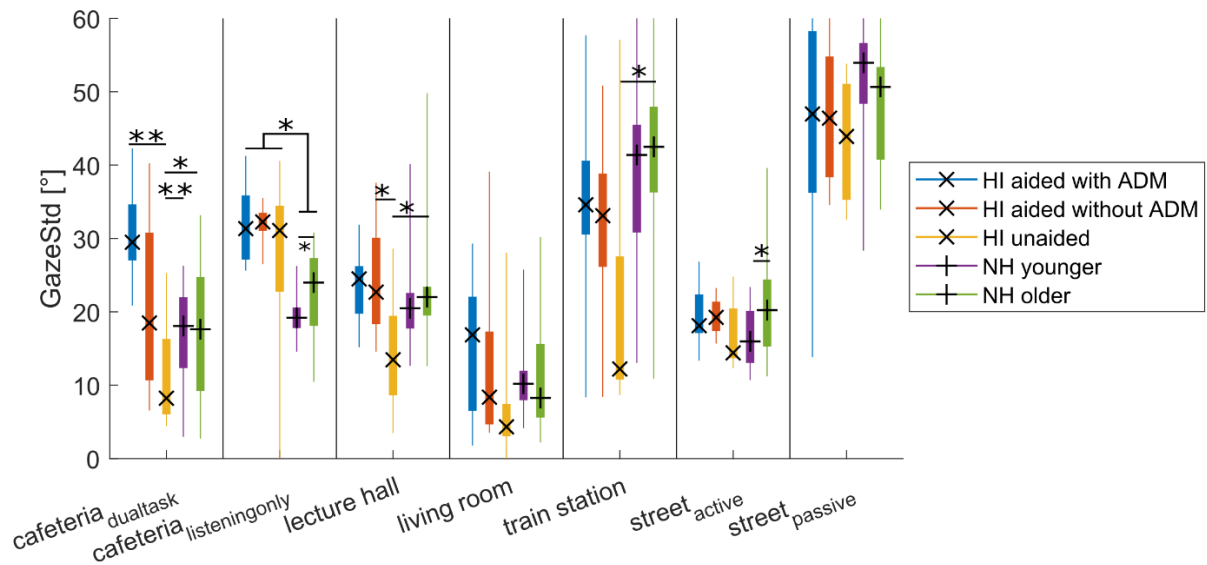

**Figure 3: Boxplots of the GazeStd for all participant groups in all VEs. Significant differences are indicated at the *.05 level, **.01 level. Boxplots show the range (thin line), 25th and 75th percentiles (thick line) and the median (cross).**

## Acoustic Simulations

The outcomes of the multiway ANOVA that was performed on the SNR measures calculated from the acoustic simulations are displayed in Table 3. There is a strong effect of the participant group on the SNRrelative, and a weak effect of the participant group on the SNRimprovementADM. Furthermore, there are medium sized interaction effects of participant group and environment on the SNRrelative and SNRimprovementADM. The SNRrelative and SNRimprovementADM are plotted in Figure 5 per participant group and VE. Because the interaction effect is significant for both measures, pairwise comparisons (with Bonferroni correction) were done between participant groups in each VE, and significant differences are indicated.

Overall, the HI participants had a significantly higher SNRrelative than the NH participants (p<.001). This occurred in the cafeteria and lecture hall VEs. Moreover, the unaided HI participants had a higher SNRrelative than the other HI participants (p<.05). This occurred in the cafeteria$_{dualtask}$ VE, where the head movements of the unaided HI participants increased the SNR by 2.2 dB. The SNRrelative of the HI participants was on average significantly higher than zero (95% confidence interval does not include zero), on average 0.6 dB. This indicates that the head movements of the HI participants increased the SNR compared to the situation without head movement (always facing frontal direction), and compared to the movements of the NH participants. The SNRrelative of the NH participants was on average significantly lower than zero, by 0.2 dB.

There were also significant differences in SNRimprovementADM between participant groups, mainly in the cafeteria$_{dualtask}$, lecture hall, and train stations VEs. All HI participants had a significantly higher SNRimprovementADM than the NH participants in the lecture hall VE, and both aided HI participant groups had a significantly higher SNRimprovementADM than the NH participants in the cafeteria$_{dualtask}$ VE. In the train station VE, the SNRimprovementADM of the younger NH participants was significantly higher than that of the HI participants. However, no significant differences in SNRimprovementADM were found between the head movement trajectories of the HI participants that were aided with the ADM, and the HI participants that were aided without ADM. The unaided HI participants had a significantly lower SNRimprovementADM than the HI participants aided with ADM in the cafeteria$_{dualtask}$ VE, but there were no significant differences in the other VEs. The SNRimprovementADM was only significantly higher than zero for the HI participants in the lecture hall (0.4 dB) and living room (1.3 dB) VEs, and for the NH participants in the living room (1.3 dB) and train station (0.3 dB) VEs.

**Table 3: Main effects of multiway ANOVA for the SNR measures.**

| Factor | Measure | F-value | p | p$_{adj}$ | Effect Size |
|---|---|---|---|---|---|
| group | SNRrelative | F(4,65)=41.9 | <.001 | <.001*** | 0.72 |
| | SNRimprovementADM | F(4,65)=4.6 | .003 | .005** | 0.22 |
| group* environment | SNRrelative | F(13.1,212.2)=19.3 | <.001 | <.001*** | 0.54 |
| | SNRimprovementADM | F(13.3,215.8)=12.5 | <.001 | <.001*** | 0.43 |

*Note.* The p$_{adj}$-column lists p-values adjusted for multiple comparisons (Benjamini-Hochberg procedure). Significances are indicated at the *.05 level **.01 level ***.001 level.

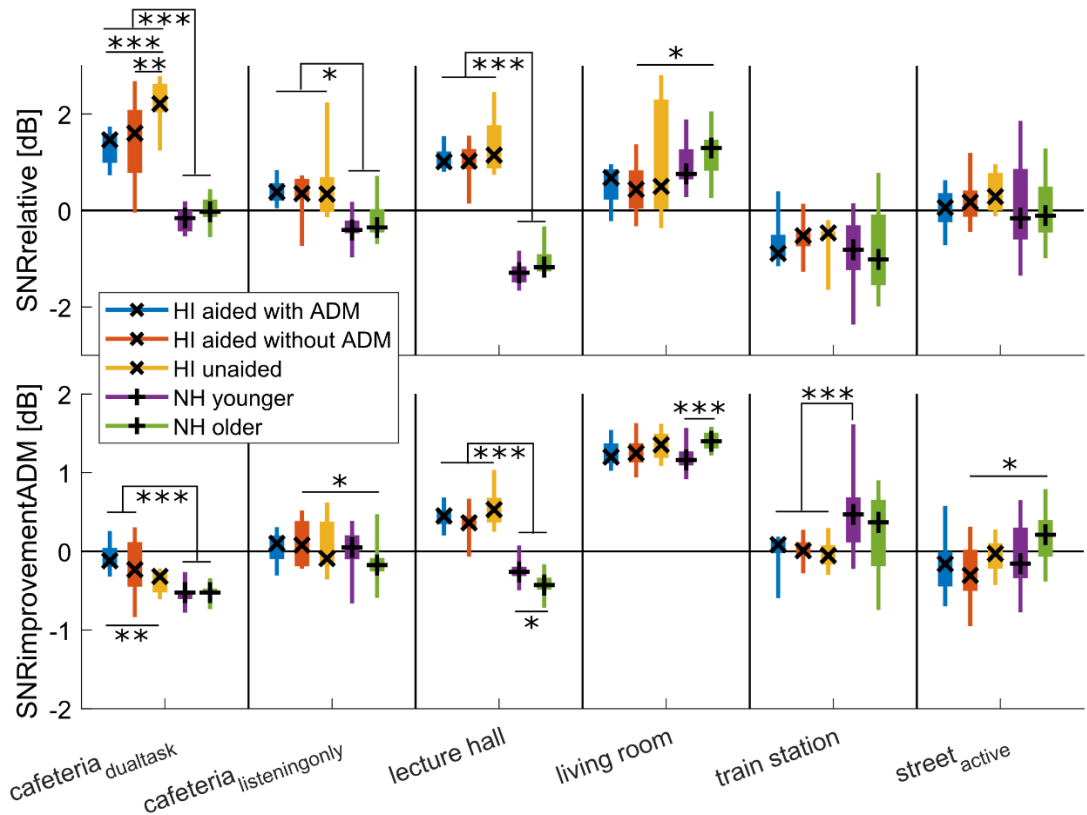

**Figure 4: The SNR resulting from the head movement of the participants, relative to the SNR without movement (always facing frontal direction), for all participant groups in all environments (top). The SNR improvement provided by the ADM for the head movement of the participants (bottom). Significant differences are indicated at the *.05 level, **.01 level, ***.001 level. Boxplots show the range (thin line), 25th and 75th percentiles (thick line) and the median (cross).**

# Discussion

The analysis of self-motion shows that, as expected, the HI participants looked at the target speaker, as did the NH participants. According to our expectations, the HI participants moved their head differently, as evidenced by the significantly higher HeadGazeRatio. This confirms the trend of an increased HeadGazeRatio that was found in NH older participants in previous work (Hendrikse et al., 2019a), but hearing impairment seems to have a much stronger effect than age. As noted in previous work, wearing glasses could also affect the HeadGazeRatio. Because all HI participants were wearing glasses this could not be checked, but it could be a confounding factor. However, surprisingly, the increased HeadGazeRatio of the HI participants did not result in a more accurate orientation of the head towards the target source (no effect of participant group on TargetHeadRMS). As shown by Brimijoin et al. (2012), pointing the head exactly at the target source may not be the optimal orientation to increase the (better-ear) SNR, but, depending on the position of the noise source(s), orienting somewhat off-axis may be better. So, it could still have been that the self-motion of the HI participants would have resulted in an increased SNR, and, as the acoustic simulations show, this was indeed the case. The SNRs resulting from the self-motion of the HI participants were significantly higher than the SNR that would have resulted from always pointing the head in the frontal direction, by 0.6 dB on average. On the contrary, the head movements of the NH participants decreased the SNR by 0.2 dB on average compared to always pointing the head in the frontal direction. These differences between HI and NH participants occurred in the cafeteria and lecture hall VEs, probably because the difference in HeadGazeRatio is only apparent when the head is oriented in more lateral directions, such as towards the off-center talkers in the cafeteria or the screen in the lecture hall.

The different groups of HI participants had a very similar self-motion. The unaided participants stand out a bit, as they moved their gaze slightly less (but only significantly less than the NH older participants), as indicated by the GazeStd. Since the self-motion was similar, the aided participants with ADM did not seem to have adapted their self-motion to increase the SNR improvement of the ADM. To investigate this further, the SNR improvement of the ADM that would have resulted from the head movement of the participants was calculated for all participant groups. The SNRimprovementADM for the aided participants with ADM was indeed not significantly different from that for the aided participants without ADM. The unaided participants had a significantly lower SNRimprovementADM in the cafeteria$_{dualtask}$ VE compared to the aided participants, which could be related to the significantly higher SNRrelative in that VE. Compared to the NH participants, the HI participants had a significantly higher SNRimprovementADM in the cafeteria$_{dualtask}$ (except the unaided HI participants) and lecture hall VEs, probably because of the differences in HeadGazeRatio. This difference between HI and NH participants is related to hearing impairment, age and/or wearing glasses, and not a result of adapting the self-motion because of the ADM. The ADM only provided a significant benefit in the living room VE (1.3 dB), in the lecture hall VE (0.4 dB) for the HI participants, and in the train station VE (0.3 dB) for the NH participants. Such a low benefit of the ADM may not be enough of an incentive to adapt the self-motion.

The aided participants with ADM had a directional mode in their own devices, and they could briefly get used to the VE and listening to the ADM output before the measurement started. However, it could be discussed that more time would be needed to get accustomed with the ADM before people adapt their self-motion. Such a study design was not yet possible, because the 'hearing aid' was implemented using the openMHA on a PC. Recent developments in hardware have resulted in a portable device running the openMHA (Tobias Herzke et al., 2018; Pavlovic et al., 2019), so it would now be possible to have a more longitudinal study design where the participants can get accustomed to the HA algorithm(s) at home. It would then also be possible, and interesting, to measure the effect of other algorithms, for example a narrower beamformer, on self-motion. Such an algorithm would be different from what participants are used to with their own devices, and would definitely require a longer time period to get accustomed to. A narrower beamformer may affect head movement more, because it has a stronger attenuation of lateral sound sources. Moreover, it potentially provides a higher benefit, which would be a stronger incentive to actually adapt the self-motion.

Finally, an additional experiment was done to check the effect of wearing headphones on self-motion and of the channel order error in the preliminary HARIRs that were used during the measurements on self-motion (see Appendix 2). The results show that wearing headphones and the channel order error in the preliminary HARIRs did not have a significant effect on the movement measures that were calculated in this study. These effects are therefore negligible and the comparison between the data of the NH participants, measured with loudspeakers, and the data of the HI participants, measured with headphones, is valid.

# Conclusion

A strong effect of hearing impairment was found on the relative amount of movement the participants did with their head and eyes (head-eye ratio). The HI participants did almost all of the movement with their head, whereas the NH participants did relatively more of the movement with their eyes. Since all HI participants were wearing glasses, this could be a confounding factor, but there was no way to check this. The different groups of HI participants had a very similar self-motion, but the unaided participants moved their gaze slightly less. Surprisingly, the increased head-eye ratio of the HI participants compared to the NH participants overall did not result in a more accurate orientation of the head towards the target direction. However, the self-motion of the HI participants did result in an increased SNR, which was estimated using acoustic simulations, both in comparison with the NH participants (0.8 dB difference) and in comparison with the situation when the head was always facing the frontal direction (0.6 dB difference). Moreover, it led to differences in the SNR improvement that the ADM would have provided in the virtual environments for the head movement trajectories of the participants.

This SNR improvement by the ADM was estimated using acoustic simulations, also for movement trajectories of the participants who were not aided with the ADM. The HI participants had a significantly higher estimated SNR improvement than the NH participants in the cafeteria$_{dualtask}$ and lecture hall VEs, which is probably related to the differences in head-eye ratio. The estimated SNR improvement was not different for the movement trajectories of the participants aided with ADM, compared to the HI participants that were not aided with the ADM during the measurements. This is in line with the finding that the self-motion was very similar for the HI participant groups. It can be concluded that the participants aided with ADM did not adapt their self-motion to increase the benefit of the ADM. The low benefit of the ADM may not have been enough incentive to adapt the self-motion and it would be interesting to test the interaction of different algorithms, for example narrower beamformers, with self-motion in the future. A narrower beamformer may affect head movement more, because it has a stronger attenuation of lateral sound sources. Moreover, it potentially provides a higher benefit, which would be a stronger incentive to actually adapt the self-motion.

# Acknowledgements

This study was funded by the Deutsche Forschungsgemeinschaft (DFG, German Research Foundation) – Project-ID 352015383 – SFB 1330 B1, and the European Regional Development Fund [ERDF-Project "Innovation network for integrated, binaural hearing system technology (VIBHear)"].

# Appendix

## Appendix 1: Descriptions of virtual audiovisual environments

A brief description of the virtual audiovisual environments is given here, as well as the presentation levels measured when presenting the virtual environments over loudspeakers. The noise level listed here is the level that was used in the previous study with NH participants. For the HI participants in the current study, the noise level was individually adjusted based on the SRT70 measurement, resulting in a noise level reduction of 6.8 ± 2.5 dB on average. Further acoustic features can be found in Hendrikse et al. (2019a).

In the living room VE, the participant sat on a sofa in the virtual living room with a person eating chips (-90°, negative angles to the right, 0° in frontal direction) next to him. A person commenting on the news (45°) sat on an armchair, and in the background a fire crackled. The participants were asked to attend to the newsreader on the TV set in front (-4°). Through the open door to the kitchen, noises of the dishwasher, water cooker and fridge could be faintly heard. Target sound level $L_{eq}$: 60.7 dBA, noise sound level $L_{eq}$: 52.5 dBA.

In the lecture hall VE, the participant sat in the audience with many animated persons. These occasionally produced distracting noises, such as coughing or writing, and the ventilation of the beamer could also be heard. The lecturer stood frontally (-15°) to the participant. Next to him a screen (25°) was placed on which the slides of the lecture about a toolbox for the creation of acoustic scenes were shown. The speech of the lecturer was reproduced by two loudspeakers (51° and -38°). During the course of the lecture a paper airplane flew over the heads of the audience and landed next to the lecturer. Target sound level $L_{eq}$: 51.7 dBA, noise sound level $L_{eq}$: 44.6 dBA.

The cafeteria VE was modelled after the cafeteria at the Carl von Ossietzky University Oldenburg Campus Wechloy. Here, the participant sat at a table with four animated persons. Three of these persons (-28°, 8°, 34°) talked to each other. In the background was cafeteria noise (diffuse recording of the Wechloy cafeteria, containing e.g. fragments of speech and plate clanging; Grimm & Hohmann, 2019), as well as music, laughter and conversations at the neighboring table. Target sound level $L_{eq}$: 63.6 dBA, noise sound level $L_{eq}$: 60.1 dBA.

The train station VE represented the Oldenburg central station. The participant stood on a platform and was asked to listen to the loudspeaker announcements, presented by several virtual loudspeakers, about trains arriving and departing. Meanwhile, trains were entering the station and animated persons were passing the participant with trolleys. On the neighboring platform four people were talking to each other (98°). In addition, the beeping of a ticketing machine could be heard. Diffuse background noise (diffuse recording of Oldenburg central station) was present. Target sound level $L_{eq}$: 68.3 dBA, noise sound level $L_{eq}$: 67.8 dBA.

The street VE is another complex multi-talker situation. The noise here consisted of moving sound sources. The participant stood at a bus stop with four animated persons (-17°, 4°, 23°, 42°). Again the task was to follow the conversation (hereafter: street$_{active}$ VE). Meanwhile, vehicles drove past the participant on the right side. The moving sound sources consisted of trucks, cars, an ambulance with sounding siren, bicycles, talking passers-by and a singing mother with a baby carriage. In addition, children were playing in a schoolyard in the background and a train was passing by. As diffuse sound sources, birdsong and street noise were part of the scene (Grimm & Hohmann, 2019). Target sound level $L_{eq}$: 63.2 dBA, noise sound level $L_{eq}$: 63.4 dBA. In addition to the street$_{active}$ VE, a measurement with passive listening in a street VE (hereafter: street$_{passive}$ VE) was carried out. Here, the participant stood at an intersection and had the task of waiting for a person he had an appointment with to arrive. The background noises were identical to those of the street$_{active}$ VE. At the end of the scene, an animated person approached the participant.

# Appendix 2: Influence of wearing headphones and HARIR channel order error

## Method

An additional small experiment was conducted to test the influence of wearing headphones on self-motion, and the influence of the error in channel order for the preliminary HARIRs. For this experiment, 9 self-reported normal-hearing (NH) participants were recruited. Due to limitations because of COVID-19, only the authors and members of the research group "Auditory Signal Processing and Hearing Devices" were available as participants for this additional experiment. The task of the NH participants was to listen to the same VEs as the HI participants, once while the audio was played back over loudspeakers and twice while the audio was played back over headphones (using preliminary and final HARIRs). For playback over loudspeakers the aforementioned setup of 29 loudspeakers was used. For the NH participants there was no individual adjustment of the noise level. The same movement measures as explained before were calculated on the head and gaze movement trajectories of the participants in this additional experiment, and a multiway ANOVA was done to compare the three listening conditions (headphones preliminary HARIRs, headphones final HARIRs, loudspeakers). As before, the false discovery rate was controlled with the Benjamini-Hochberg procedure (Benjamini & Hochberg, 1995).

## Results

The outcomes of the multiway ANOVA comparing the listening conditions are shown in Table 4. The results show that there was no significant effect of listening condition on any of the movement measures. This indicates that the effects of wearing headphones and channel order error are negligible.

**Table 4: Main effects of the multiway ANOVA to compare listening conditions.**

| Factor | Measure | F-value | p | $p_{adj}$ | Effect Size |
|---|---|---|---|---|---|
| listening condition | GazeStd | $F(2,14)=3.1$ | .076 | .608 | 0.31 |
| | HeadStd | $F(1.2,8.1)=1.1$ | .342 | .684 | 0.13 |
| | GazeSpeedMean | $F(2,14)=1.5$ | .248 | .992 | 0.18 |
| | HeadSpeedMean | $F(2,14)=1.4$ | .291 | .776 | 0.16 |
| | NGazeJumps | $F(2,14)=1.0$ | .403 | .645 | 0.12 |
| | HeadGazeRatio | $F(2,14)=0.7$ | .504 | .672 | 0.09 |
| | TargetGazeRMS | $F(1.1,7.6)=0.4$ | .590 | .674 | 0.05 |
| | TargetHeadRMS | $F(2,14)=0.4$ | .693 | .693 | 0.05 |

*Note.* The $p_{adj}$-column lists p-values adjusted for multiple comparisons (Benjamini-Hochberg procedure). Significances are indicated at the *.05 level **.01 level ***.001 level.